# Si Orientation Dependent Release of SiO$_2$ Microcantilevers by Wet Chemical Etching Method

Sagnik Middya[1], K. Prabakar[2]

[1]Department of Electronics and Electrical Engineering,
Indian Institute of Technology, Guwahati - 781039, Assam.

[2]Electronics and Instrumentation Section, Surface and Nanoscience Division,
Indira Gandhi Centre for Atomic Research, HNBI, Kalpakkam-603102, Tamil Nadu.

**Abstract**
Micro-cantilevers (MCs) are widely used in sensing, probing and a variety of other purposes. Apart from Si, the most versatile material in the "micro" domain, SiO$_2$ MCs have attracted much attention lately. In this study, fabrication of SiO$_2$ MCs using wet anisotropic etching technique is reported. Thermally grown SiO$_2$ (300 nm) layer on single crystalline($c$-) Si is utilized for the micro-fabrication. Standard lithographic procedure was followed and the exposure of the photo-resist was performed by mask-less direct laser writer. Various orientations on the $c$-Si wafer, their properties and implications on the structure and integrity of the released SiO$_2$ MC are discussed. The <110> direction was found to be much time consuming and sticking of the released structures was almost evident. On the other hand <100> and <410> directions showed lesser release time and higher stability of the released structures due to under-etching along the width of the MCs.

## 1. Introduction

Use of MCs as a mechanical device was actually initiated by Nathanson et al. [1] when they used metal plated MCs as the gate electrode of resonant gate transistor. A variety of aspects like bending, vibration, torsional modes of MCs can be utilized as transduction principle in a wide range of innovative sensing devices [2]. Surface probe microscopy (SPM) is by far the best example of its application. By the virtue of its well established micromachining techniques and other unique properties, Si stands out to be the most suitable candidate for MCs at a length scale of few tens or at most hundreds of μm. Apart from surface probing, these cantilevers are often used as chemical (in itself or by using a layer of some specific material), [3,4,5] pressure [6] or thermal (utilizing coefficient of expansion) [7] sensors. Piezo-electric or other actuation principles coupled with MCs makes them high end actuators.

Similar principle can be extended to develop energy harvesters [8]. Several other possible applications of MCs in the form of heaters [9], humidity sensors have been reported [10]. In the past years, SiO$_2$ has been found to be an alternative for Si and many SiO$_2$ based MC systems have been fabricated for a variety of applications [11,12]. Simulations show that for the same load, SiO$_2$ MCs register a greater bending as compared to Si MC [13]. Hence SiO$_2$ MCs prove to be very useful for many practical purposes.

In this work, we present the results of Si crystalline orientation dependent fabrication of SiO$_2$ MCs by direct laser writer and wet anisotropic chemical etching process.

## 2. Experimental section

Experiments were conducted on 2 x 2cm cut, $c$-Si(100) wafers with 300 nm thermally grown SiO$_2$ on both the sides. Fabrication started with standard Si wafer cleaning using standard Radio Corporation of America developed RCA1, RCA2 and *Piranha* solution (3:1 mixture of $H_2O_2$ to $H_2SO_4$). The wafer pieces were rinsed in acetone $((CH_3)_2CO)$ and isopropyl alcohol $((CH_3)_2CHOH)$ to remove any organic contaminants after it was brought into clean room. It was placed on hot plate at 100°C for a minute to remove the moisture. Then positive S1818 photoresist was spin coated. Thickness of 1 μm and was confirmed by non contact profiler (Dektak) measurement. Subsequently, it was heated at 65°C (pre-baking) for about 1 minute to evaporate the solvents in photoresist.

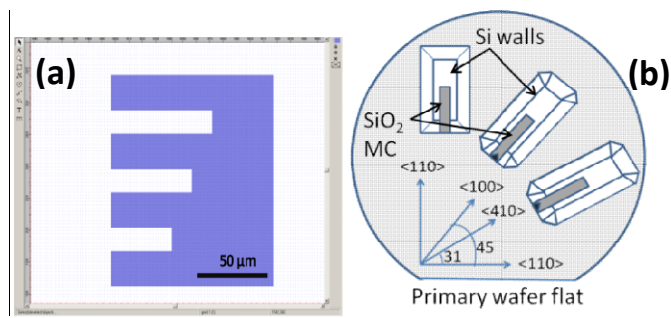

FIG 1: (a) Design of microcantilever array in CleWin Software (b) Different crystallographic orientations and microcantilevers in Si <100> wafer with respect to primary flat.

A design consisting of individual MCs and MC arrays of various dimensions were created using CLEWIN software. A particular design of one such MC array is shown in figure 1(a). The MCs (represented by white projections in the figure) had dimensions of 75 μm, 60 μm and 45 μm. The pattern consists of MCs aligned to three crystallographic orientations of Si namely <100>, <110> and <410>. A schematic representation of these three directions on a <100> Si wafer with respect to the primary wafer flat is illustrated in figure 1(b).

Usually in photolithography, the photoresist is flood-exposed by UV light through a mask containing the pattern. The exposed photoresist region either becomes soluble (positive photoresist) or insoluble (negative photoresist). Here a mask-less technique was adopted. The patterns were transferred onto the photoresist coated wafer using a Direct Laser Writer (LW 405B, M/s Microtech, Italy). The instrument raster-scans a laser beam (wavelength 405 nm) with variable spot size. There is also a provision for auto-focusing the laser beam before executing each scan. Suitable laser power and the patterns are selected from the user console. The spot size can be varied by choosing between different lenses. In our case, a laser power of around 10 mW and spot size of 2 μm was chosen. After exposure, the pattern was developed in CF60 developer for about 45 seconds followed by an immediate dip in deionized water. Every pattern developed was observed under microscope and if found under developed, was kept in developer for few more seconds or if over developed, photo resist was stripped and lithography was repeated. Once the patterns were satisfactory, the wafer was post baked at 120°C for 15 minutes for hardening of the photoresist.

The pattern was transferred from photoresist to wafer, by etching $SiO_2$ using buffered HF (100 gm of $NH_4F$, 150 ml of DI water (V) with (V/3) of HF). This gives etch rate of about 300 nm/min. Complete removal of the oxide can be noticed when the hydrophilic oxide turns to hydrophobic nature of the underlying Si. The residual photoresist was stripped in acetone and isopropyl alcohol. Si exposed through the $SiO_2$ hard mask was etched in the present work to release the $SiO_2$ cantilever using KOH (potassium hydroxide). KOH is an anisotropic etchant of Si. Selectivity of KOH etching to crystallographic orientation is <100> : <111>: 66:1 and selectivity against oxide is greater than 200. In the present experiment, 40 % aqueous KOH was chosen as the etching solution. The solution was taken in non interacting alumina crucible and was heated to 75°C (etch rate of 45 μm/h). After the etching process, the etched wafers were removed from the solution and thoroughly rinsed with deionized water without the application of any procedure for the prevention of cracking or stiction of the released MCs. The released MCs were allowed to dry in ambient air and were characterized using optical microscope.

## 3. Results and discussions

The crystallographic orientation of Si plays a major role while releasing the $SiO_2$ MCs using chemical wet etching. There are two important aspects, etch time and stability of the released MC which need to be considered. It is well known that Si has a diamond cubic structure and its anisotropic etching behaviour strongly depends on the crystal orientation. For example, the {111} planes are most resistant to etching due to high density of atoms and hence in majority of the cases they form the boundary for the etched structure. To achieve {111} walls, the pattern edges must be in the <110> direction, which is also the direction of the prime flat in Si <100> wafers. The family of {100} planes in Si posses fourfold symmetry and their anisotropic etching can produce either vertical {100} walls or sloping {110} or {111} walls, inclined at 45° or 54.7°, respectively [14]. Similarly, the sectional line of a {411} and a {111} plane points in the <410> direction, forming an angle of 30.96° with <110> direction. These orientations and their relative directions are shown in figure 1(b). These orientations determine the way etching of Si below the MCs progress which in turn impacts their properties.

Figure 2 illustrates an optical image of <110> oriented partially released MCs (parallel to primary wafer flat). Bound by slow etching {111} planes, it is difficult to etch below the MCs along these directions. Etching starts once some fast-etching planes are exposed at the convex corners and progresses along the length of the MC [14], known as frontal under-etching. As an outcome, releasing times of MCs are dependent on their length and the structures tend to be unstable. In this process there is also a possibility that the MCs bend and get stuck to the floor of the etch-pit. Figure 2 depicts the progress of under-etching as well as the length dependency of etch-time. While the smallest MC has been released completely, the others are yet to be released. It may be noted all the MCs released in the present work, appear transparent under optical microscope because of the ~300 nm thickness of $SiO_2$.

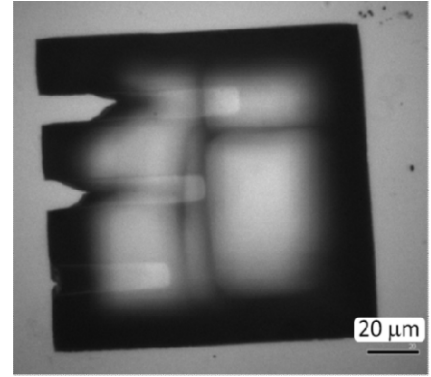

FIG 2: Optical image of array of Microcantilevers along <110>. The $SiO_2$ MCs appear transparent. The smallest MC has been released while the others are yet to be released.

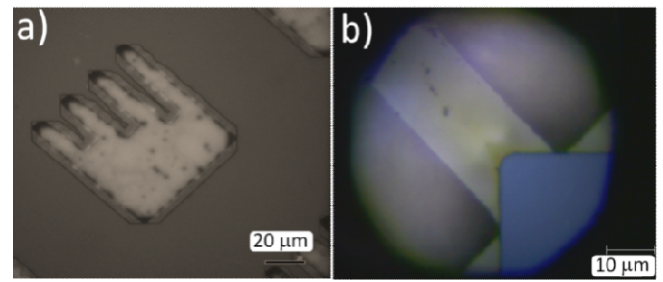

FIG 3: Optical images of <100> oriented MC a) yet to be completely released microcantilevers and b) showing symmetrical triangular projection of {111} Si planes at the fixed end of the completely released microcantilevers.

In the <100> direction, both {100} and {110} planes are present. If KOH is chosen as the anisotropic etchant as in our case, the Si projections below the unreleased MCs are bounded by {100} planes resulting in vertical walls. This can be justified by considering that {110} planes etch faster than their {100} counterparts in KOH solution. Both the {100} and {110} planes can be etched quickly and hence under-etching proceeds laterally i.e. along the width of the MCs as indicated in figure 3(a). As a consequence, MCs require less time to be released. The release-time depends on the width and is almost independent of the lengths of the MCs. As under-etching proceeds along the width, the gradually thinning line of Si acts as a support for the MC thus preventing it from collapsing [14]. Figure 3(b) shows that even on complete release, there is still a symmetrical triangular projection of {111} Si planes at the fixed end of the MCs. This happens because {111} planes are relatively etch-resistant.

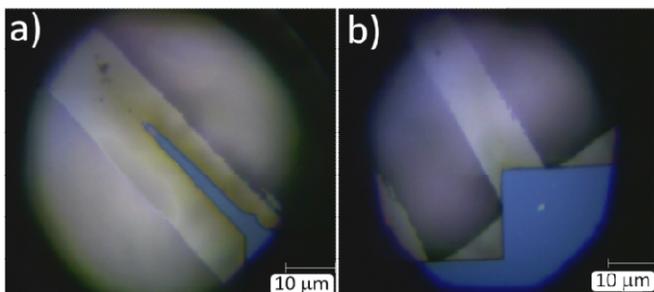

FIG 4: Optical images of <410> oriented MC a) yet to be released and b) completely released and showing unsymmetrical triangular projection of {111} Si planes at the fixed end of the MC

As mentioned earlier, the fast-etching {411} plane intersect the {100} plane (also the wafer surface) along the <410> direction inclined at 31° with respect to the primary flat. Thus, the MCs oriented along this direction are released by under-etching of {411} planes. This direction also exhibits some sort of lateral under-etching (Figure 4(a)) though not uniform on both sides as in case of <100> direction. Much like <100> direction, here we also observe a triangular projection of {111} Si planes at the fixed end of the released MCs. However, as figure 4(b) illustrates, in this case it is unsymmetrical. As mentioned earlier in the context of <100> direction, this lateral under-etching feature has the advantage that the released MCs are stable. A comparison between various MCs released in the present work is given in table 1.

As a summary, in the present study, SiO$_2$ micro-cantilevers of various dimensions and crystallographic orientations were fabricated on Si by wet chemical etching method. The released MCs where characterized using optical microscope. It is found that the release time of MCs oriented in the direction of <100> are dependent on their length due to the frontal etching. Hence for an array consisting of different dimensions of MCs, at any point of time the already released MCs have a greater chance of collapsing or sticking with the bottom.

TABLE 1: Orientation of Si wafer and properties of the released SiO$_2$ MCs along those directions. Angle is with respect to primary flat of the Si <100> wafer used.

| Direction | Angle | Remarks on released MCs |
|---|---|---|
| <110> | 0° | 1. Frontal etching<br>2. Release time depends on MC length<br>3. MCs are unstable<br>4. Shortest MCs are completely released. Some MCs tend to stick to the bottom |
| <100> | 45° | 1. Lateral etching<br>2. Release time is independent of MC length but depends on its width<br>3. MCs are stable<br>4. Though the MCs are released, there is a symmetrical triangular projection of Si below the fixed end of MCs |
| <410> | 31° | 1. Lateral etching<br>2. Release time is independent of MC length<br>3. MCs are stable<br>4. MCs are completely released, but unlike <100>, the projections of Si below the fixed end of MCs are unsymmetrical |

Whereas, in the case of MCs oriented in the direction of <110> and <410>, the release time was independent of its length because of lateral etching. Quite intuitively these structures are more stable as a thin line of Si remains below the MC throughout the vigorous etching process until it gets dissolved. It can be concluded that in spite of the remnant triangular Si projection at the fixed end of the MCs, the <110> and <410> direction shows better resilience to fabrication.

## Acknowledgements

Authors thank Mr. Raghuramaiah, EIS, MSG, IGCAR for useful discussions.